\documentclass[12pt,preprint]{aastex}
\lefthead{Skinner et al.}
\righthead{Westerlund 1}

\begin{document}

\title{A Rich Population of X-ray Emitting Wolf-Rayet Stars in
       the Galactic Starburst Cluster Westerlund 1}

\author{S.L. Skinner, A.E. Simmons}
\affil{CASA, Univ. of Colorado, Boulder, CO 80309-0389 }

\author{S.A. Zhekov}
\affil{Space Research Institute, Moskovska str. 6, Sofia-1000, Bulgaria}

\author{M. Teodoro, A. Damineli}
\affil{Instituto de Astronomia, Geof\'{i}sica e Ci\^{e}ncias Atmosf\'{e}ricas,
       Universidade de S\~{a}o Paulo, Rua do Mat\~{a}o 1226, Cidade 
       Universit\'{a}ria, 05508-900 S\~{a}o Paulo, SP, Brazil}

\author{F. Palla}
\affil{INAF-Osservatorio  Astrofisico di Arcetri, Largo E. Fermi 5, 50125 Firenze, Italy}

%
\newcommand{\ltsimeq}{\raisebox{-0.6ex}{$\,\stackrel{\raisebox{-.2ex}%
{$\textstyle<$}}{\sim}\,$}}
%
\newcommand{\gtsimeq}{\raisebox{-0.6ex}{$\,\stackrel{\raisebox{-.2ex}%
{$\textstyle>$}}{\sim}\,$}}

\begin{abstract}
Recent optical and infrared studies have revealed that the heavily-reddened 
starburst cluster  Westerlund 1 (Wd 1) contains at least 22 Wolf-Rayet (WR)
stars, comprising the richest WR population of any galactic cluster.
We present results of a sensitive {\em Chandra} X-ray observation of 
Wd 1 which  detected 12 of the 22 known WR stars and the mysterious 
emission-line star W9.  The fraction of detected WN stars is nearly
identical to that of WC stars. The WN stars WR-A and WR-B as well as 
W9 are exceptionally luminous in X-rays and have similar hard 
heavily-absorbed X-ray spectra with strong Si XIII  and 
S XV  emission lines. The luminous high-temperature X-ray
emission of these three stars is characteristic of colliding wind
binary systems but their binary status remains to be determined.
Spectral fits of the X-ray bright sources WR-A and W9 with isothermal 
plane-parallel shock models require high absorption column densities
log N$_{\rm H}$ = 22.56 (cm$^{-2}$)  and yield characteristic shock
temperatures  kT$_{s}$ $\approx$ 3 keV 
(T$_{s}$ $\approx$ 35 MK).
\end{abstract}



%
\newpage

\section{Introduction}
The heavily-reddened  cluster Westerlund 1 (Wd 1; Westerlund 1961, 1987)
in Ara has recently been recognized as a rare example of a starburst
cluster in our own Galaxy.  The cluster is massive, compact, and young 
with age estimates of $\sim$3 - 5 Myr (Brandner et al. 2005 = B05;
Clark et al. 2005 = C05). Wd 1  contains
a remarkable collection of massive post-main sequence stars including
early and late-type supergiants, a luminous blue variable candidate,
and the largest known population of  Wolf-Rayet (WR) stars of any
galactic cluster (C05; Negueruela \& Clark 2005 = NC05; 
Clark \& Negueruela 2002). A recent {\em VLT} study has also revealed
a faint population of  low-mass pre-main sequence stars and gives a
photometric distance d = 4.0 $\pm$ 0.3 kpc (B05), but spectroscopic
studies allow a larger range of distances (C05). The extinction is 
A$_{\rm V}$ $\approx$ 9.5 - 13.6 mag (B05, C05). 

Ongoing studies have so far identified 22 WR stars in Wd 1 and the census
is likely incomplete. WR stars are highly-evolved  evolutionary  descendants 
of massive O-type stars that are in advanced nuclear burning stages and
undergoing extreme mass-loss from high-velocity winds, rapidly approaching
the end of their lives as supernovae. At least one supernova has already
occurred in Wd 1 as evidenced by the  discovery of a new X-ray pulsar
(Figure 1; Skinner et al. 2005a = S05a; Muno et al. 2006).

There is at present no  comprehensive theory of X-ray emission from WR stars.
Previous observations have focused mainly on X-ray bright  WR $+$ OB binaries such
as $\gamma^2$ Vel and WR 140 (Skinner et al. 2001; Zhekov \& Skinner 2000),
whose hard emission  (kT $\geq$ 2 keV) is thought to originate primarily
in colliding wind shocks. Much less is known about the X-ray emission of
single WR stars, but by analogy with O-type stars they are expected to emit 
soft X-rays (kT $<$ 1 keV) from instability-driven shocks formed in their 
supersonic winds (Gayley \& Owocki 1995). Despite these expectations, X-ray 
emission from single WR stars has proven difficult to detect. Recent sensitive
observations have shown that single carbon-rich WC stars are exceedingly
faint in X-rays or perhaps even X-ray quiet,for reasons that are not yet 
fully understood (Skinner et al. 2005b = S05b). Additional X-ray observations
are needed to quantify X-ray emission properties across the full range of
WC and WN spectral subtypes.

The presence of a rich, equidistant, coeval population of WR stars in Wd 1
makes it an opportune target for X-ray observations, which are capable of
penetrating the high extinction.
We present the results of a sensitive {\em Chandra} X-ray observation of
Wd 1, focusing here on the WR population. This observation yields 12 new
WR X-ray detections and provides valuable new information on the X-ray
properties of this unique sample of galactic WR stars that can be used
to test shock emission models and guide new theoretical development.

\section{Chandra Observations}

{\em Chandra} observed Wd 1 on 22-23 May 2005 
and 18-19 June 2005 with exposure live times
of  18,808 s and 38,473 s respectively.  
The observations were obtained with  the 
ACIS-S imaging array in timed faint event mode 
using a 3.2 s frame time. 
The pointing positions were (J2000.0): 
R.A. = 16$^{\rm h}$ 47$^{\rm m}$ 08.60$^{\rm s}$, 
$-$45$^{\circ}$ 50$'$ 27.4$''$  in
May 2005  and 
R.A. = 16$^{\rm h}$ 47$^{\rm m}$ 07.78$^{\rm s}$, 
$-$45$^{\circ}$ 51$'$ 00.9$''$ in June 2005.

Data reduction was based on Level 2 event files
generated  by the {\em Chandra} X-ray Center.
Source detection
was accomplished using the CIAO
\footnote{Further information on {\em Chandra} Interactive
Analysis of Observations (CIAO) software can be found at
http://asc.harvard.edu/ciao.}  (vers. 3.2.1)
tool {\em wavdetect} applied to full resolution
images (0.$''$492 pixel size) that were energy filtered
to include only events in the [0.3 - 7] keV energy range
to reduce background.  The 3$\sigma$
elliptical source regions generated by {\em wavdetect}
were used to extract an event list for each source.
The source event lists were used for further timing
and spectral analysis. The probability of constant
count rate P$_{c}$ was computed for each source using the
non-parametric K-S statistic (Skinner, Gagn\'{e},
\& Belzer 2003 and references therein). Spectra
and associated  instrument response files for brighter
sources were extracted from updated Level 2 event files
using recent  CIAO vers. 3.3 tools that incorporate
the latest gain and effective area calibrations (CALDB vers. 3.2).
Spectra were analyzed using XSPEC vers. 12.2.0.

\section{Wolf-Rayet Stars}

\subsection{Wolf-Rayet Star X-ray Detections}
{\em Chandra} detected 12 of the 22 known WR stars in Wd 1,
including 11 of the 19 WR stars in the list of NC05 and
1 of the 3  WR stars (all of WN subtype) identified  
by Groh et al. (2006).  Their positions are shown in Figure 1
and X-ray properties are summarized in Table 1. 
The detection rate was similar for 
WN and WC stars. Specifically, 8 of 15 (53\%) of the
known WN stars were detected and   
4 of 7 (57\%) WC stars. However, the WC9 star WR-E is 
considered to be a marginal detection. 

Three WR detections show signs of variability, but two
of these are faint sources with few counts on which to
base a variability analysis. Both  WR-K (source 2) and
WR-G (source 3) were faintly detected in the first observation
but not in the deeper second observation. WR-G had a 
low probability of constant  count rate P$_{c}$ = 0.003
in the first observation and a noticeably high mean photon energy.
WR-B (source 7) is a suspected  WN8 $+$ OB binary (NC05)
and had P$_{c}$ = 0.15 in the second observation.
A low-amplitude rise and fall can be seen in its X-ray
light curve but no variability was seen in WR-B during
the first observation.

Table 1 gives the X-ray luminosities of WR stars in Wd 1 and
the L$_{x}$ distribution is shown in Figure 2. 
The median X-ray luminosity of the 
detected WN stars  log L$_{x}$ = 32.23 (ergs s$^{-1}$) is only
slightly larger than the median log L$_{x}$ = 32.02 for
WC stars. The stars WR-A (source 13) and WR-B (source 7), 
both of which have uncertain WN spectral types (C05),
have very high L$_{x}$ and are very likely binaries.
At the other extreme, 45\% of the WR stars
in Wd 1 were undetected and the shape of the L$_{x}$ distribution
at low luminosities is not well-determined.

\vspace*{0.1cm}

\subsection{Wolf-Rayet Star Non-detections}

{\em Chandra} did not detect 10 of the 22 known WR stars in
Wd 1 down to the  detection limit log L$_{x}$ (0.3 - 7 keV) 
$\approx$ 31.3 ergs s$^{-1}$, which assumes a 6 count  threshold
in 57.3 ksec and underlying thermal spectrum with  kT = 1 keV and
N$_{\rm H}$ = 3$\times$ 10$^{22}$ cm$^{-2}$. Higher X-ray absorption
in the metal-rich winds of WC stars would  decrease the chance 
of their X-ray detection but does not explain why the 
fraction of undetected WN stars is just as high as WC stars.
If the non-detections are predominantly single stars that emit
only softer X-rays at kT $<$ 1 keV, as occurs for many O-type
stars, their emission would be preferentially absorbed and 
they could escape detection.

Although N$_{\rm H}$ and X-ray temperature  clearly affect  
X-ray detectability, convincing evidence is now emerging for very large
differences in  L$_{x}$ in WR stars with similar spectral types.
In the Wd 1 sample, the WC9 star WR-F (source 6)  was clearly detected as
a moderately bright X-ray source (log L$_{x}$ = 32.60 ergs s$^{-1}$) 
but the WC9 stars  WR-M and WR-H were 
undetected with count rate limits that are at least ten times
smaller. Furthermore, we note that a previous 20 ksec
{\em Chandra} observation failed to detect the single WC8 star WR 135 in 
Cygnus with a conservative upper limit  log L$_{x}$ (0.5 - 7 keV)
$\leq$ 29.82 ergs s$^{-1}$, which gives  a remarkably low ratio
log [L$_{x}$/L$_{bol}$] $\leq$ $-$9.1 (S05b).
Assuming  d = 1.74 kpc and low extinction A$_{\rm V}$ = 1.26 mag  
(van der Hucht 2001), it is difficult to attribute the WR 135
non-detection entirely to absorption.  
Despite their similar WC8-9 spectral types,
WR-F and WR 135 differ in  L$_{x}$ by at least a factor of 
$\approx$600. There are no indications for binarity in WR 135 and
other attempts to detect apparently single WC stars have yielded 
negative results (S05b). Thus, single WC stars emit X-rays at very
low levels (if at all) and the elevated X-ray emission of
WC stars such as WR-F in Wd 1 is very likely the result of extraneous
factors such as binarity.

\subsection{Wolf-Rayet Star X-ray Spectra}
The {\em Chandra} spectra of the brightest WR detections
reveal similar properties. They are heavily absorbed
below $\approx$1 keV and have significant emission above
kT $\simeq$ 2 keV. The spectrum of the brightest WR detection
WR-A (Fig. 3)  shows low-energy absorption as well 
as strong Si XIII (1.86 keV) and S XV (2.46 keV) emission lines.
These lines emit maximum power at log T$_{max}$ = 
7.0 (K) and 7.2 (K) respectively. The spectrum of WR-B is 
similar and  shows prominent Si XIII and S XV lines,
as does W9 (Fig. 3).

The presence of hotter plasma is not anticipated from models
of radiative shocks distributed in the winds of single stars.
The harder spectra detected here are clearly of a different
origin, and colliding wind shocks in binary systems are a 
plausible explanation. To investigate this, we have fitted
the spectrum of WR-A with the plane-parallel 1T shock model
{\em vpshock} (Borkowski, Lyerly, and Reynolds 2001) in
XSPEC vers. 12.2 using the most recent APED atomic data base
(neivers 2.0 in XSPEC). 

The {\em vpshock} model gives very good fits for WR-A with
shock temperatures kT$_{s}$ = 3.5 [2.5 - 4.8; 90\% conf.] keV,
N$_{\rm H}$ = 3.6 [3.1 - 4.2] $\times$ 10$^{22}$ cm$^{-2}$, and reduced
$\chi^2$ = 1.0 - 1.1. The above N$_{\rm H}$ equates to
A$_{\rm V}$ = 16.2 [14.0 -  18.9] mag or 
E(B-V) = 5.45 [4.70 - 6.36] using Gorenstein (1975).
Two-temperature optically thin thermal plasma models give
similar N$_{\rm H}$ values. By comparison, previous studies
of the OB supergiants in Wd 1 yield median values 
A$_{\rm V}$ = 13.6 mag or E(B-V) $\approx$ 4.35 (C05).
This suggests that the extinction across Wd 1 is quite inhomogeneous
or excess absorbing material such as cold gas is present toward
WR-A that has escaped optical detection.  The {\em vpshock}
fits give an upper limit on the ionization timescale
log $\tau$ $\leq$ 11.2 (s cm$^{-3}$), where 
$\tau$ = n$_{e}$t$_{s}$,  n$_{e}$ is the postshock electron
density, and  t$_{s}$ is the shock age. Such a low value of
$\tau$ implies that non-equilibrium ionization effects in the
shocked plasma may be important.

\section{The Unusual Emission Line Star W9}
The enigmatic  emission line star W9 lacks any recognizable
photospheric features in its R-band spectrum and shows a
very broad H$\alpha$ line (C05). It was classified as
a B[e] supergiant by C05 but its nature is uncertain and they note
that it could contain a WR component so we discuss it here.
{\em Chandra} detected a strong X-ray source (source 4 in Table 1)
at an offset of 0.$''$3 from the position of W9 given in C05. 
Two radio sources lying $\approx$15$''$ to the east of W9
identified as Ara A (N) and  Ara A (S) by Clark et al. (1998)
were located near the {\em Chandra} aimpoint but not detected.

The X-ray properties of W9 are very similar to the X-ray
bright WN star WR-A. They have nearly identical mean photon
energies (Table 1), L$_{x}$ (Fig. 2), and 
spectra (Fig. 3). There is little doubt that their X-ray 
emission is due to the same process and we suspect that 
both W9 and WR-A are colliding wind binaries. Spectral fits
of W9 with the {\em vpshock} model give values for N$_{\rm H}$,
kT$_{s}$, and $\tau$ that are within 30\% of those quoted above for
WR-A and the best-fit  W9  column density is  
N$_{\rm H}$ = 3.6 [2.6 - 4.9; 90\% conf.] $\times$ 10$^{22}$,
or A$_{\rm V}$ = 16.4 [11.8 - 22.3] mag. 
cm$^{-2}$. Thus, as for WR-A, the  X-ray absorption may exceed
that expected from previous A$_{\rm V}$ estimates.

\section{Conclusions}
There are good reasons to believe that most of
the WR X-ray detections in Wd 1 are binaries. This conclusion 
is more secure for WC stars than WN stars since there have been
no previous X-ray detections of single WC stars, even at better
sensitivities than obtained  here. Large differences in L$_{x}$ 
between WR stars of similar spectral type can be naturally
explained if the luminous X-ray sources are colliding wind 
binaries. Furthermore, {\em Chandra} preferentially detects
harder X-ray sources in Wd 1 because of the high extinction.
Plane-parallel shock models give good fits of the brightest
X-ray detections and require shock temperatures kT $\geq$ 2 keV.
Such temperatures are higher than predicted for radiative shocks
distributed in the winds of single stars, but are consistent with
colliding wind emission in binary systems. Even so, more definitive
proof of binarity is needed from optical/IR follow-up work. And,
interesting questions remain in the X-ray regime. What is the
origin of the excess absorption that is inferred from X-ray
spectral fits of WR-A  and W9? How does the WR X-ray luminosity
function behave at low L$_{x}$: are the undetected WR stars
faint sources below our detection limit or are they  X-ray quiet?

\acknowledgments

This research was supported by NASA  grants 
GO4-5003X and GO5-6009X. 

\clearpage
\begin{deluxetable}{lllllcllll}
\tabletypesize{\scriptsize}
\tablewidth{0pt}
\tablecaption{Wolf-Rayet Star X-ray Sources in Westerlund 1\tablenotemark{a} }
\tablehead{
\colhead{No.}      &
\colhead{R.A.}     &
\colhead{Decl.}    &
\colhead{Counts}   &
\colhead{Rate}     &   
\colhead{$<$E$>$}   &
\colhead{P$_{c}$}   &
\colhead{log L$_{x}$}  &
\colhead{Identification} & 
\colhead{WR Id.}               \\
\colhead{   }    &
\colhead{(J2000)     }    &
\colhead{(J2000)     }    &
\colhead{(c)            }    &
\colhead{(c s$^{-1}$)     }   &
\colhead{(keV)  }   &
\colhead{       }   &
\colhead{(ergs/s)}    &
\colhead{       }   &
\colhead{       }
}
\startdata
1\tablenotemark{c}  & 16 46 59.91 & -45 55 25.6 &  11 $\pm$  4                  & 5.85E-04 & 2.84 & 0.46    & 32.06                  & 2M 164659.90-455525 & N (WC) \nl
2\tablenotemark{c}  & 16 47 03.04 & -45 50 43.4 &   9 $\pm$  3\tablenotemark{b} & 4.82E-04 & 2.22 & 0.57    & 31.98                  & 2M 164703.15-455043 & K (WC) \nl
3\tablenotemark{c}  & 16 47 04.06 & -45 51 25.1 &  13 $\pm$  4                  & 6.93E-04 & 3.99 & 0.003   & 32.14                  & NT 164704.00-455125 & G (WN) \nl
4\tablenotemark{d}  & 16 47 04.14 & -45 50 31.4 & 334 $\pm$ 19                  & 8.68E-03 & 2.63 & 0.63    & 33.77\tablenotemark{g} & 2M 164704.15-455031 & W9\tablenotemark{e} \nl
5\tablenotemark{d}  & 16 47 04.19 & -45 51 07.2 &  67 $\pm$  9                  & 1.74E-03 & 2.82 & 0.85    & 32.54                  & NT 164704.23-455107 & L (WN) \nl
6                   & 16 47 05.21 & -45 52 25.1 &  77 $\pm$  9                  & 2.01E-03 & 3.16 & 0.49    & 32.60                  & NT 164705.23-455225 & F (WC) \nl
7\tablenotemark{d}  & 16 47 05.37 & -45 51 04.9 & 185 $\pm$ 14                  & 4.81E-03 & 2.49 & 0.15\tablenotemark{h}  & 33.57\tablenotemark{g} & NT 164705.35-455104 & B (WN) \nl
8                   & 16 47 05.99 & -45 52 08.3 &   6 $\pm$  3\tablenotemark{b} & 1.56E-04 & 1.90 & 0.61    & 31.49                  & GS 164705.99-455208 & E (WC) \nl
9                   & 16 47 06.01 & -45 50 23.1 &  27 $\pm$  6                  & 7.01E-04 & 2.96 & 0.82    & 32.14                  & 2M 164706.01-455023 & R (WN) \nl
10                  & 16 47 06.26 & -45 51 26.8 &  16 $\pm$  5                  & 4.05E-04 & 5.02 & 0.49    & 31.90                  & NT 164706.30-455126 & D (WN) \nl
11                  & 16 47 07.62 & -45 49 22.3 &  17 $\pm$  4                  & 4.29E-04 & 3.92 & 0.70    & 31.93                  & 2M 164707.61-454922 & 3 (WN)\tablenotemark{f} \nl
12                  & 16 47 07.65 & -45 52 36.0 &  40 $\pm$  7                  & 1.03E-03 & 2.30 & 0.43    & 32.31                  & 2M 164707.64-455235 & O (WN) \nl
13\tablenotemark{d} & 16 47 08.35 & -45 50 45.5 & 500 $\pm$ 23                  & 1.30E-02 & 2.68 & 0.92    & 33.92\tablenotemark{g} & NT 164708.34-455045 & A (WN) \nl

\enddata
\tablenotetext{a}{
Notes:~Data are from the 38,473 s exposure on 18-19 June 2005 unless otherwise noted.
All quantities are computed using events in the 0.3 - 7.0 keV energy range.
{\em Chandra} positions are from full-resolution (0.$''$492 pixel)  ACIS-S images.
X-ray counts inside $wavdetect$ 3$\sigma$ source detection regions are background-subtracted.
$<$E$>$ is the mean photon energy and P$_{c}$ is the probability that the count
rate was constant based on the K-S statistic. Unabsorbed X-ray luminosities L$_{x}$ 
(0.3 - 7 keV)  are from PIMMS simulations using a 1T Raymond-Smith thermal plasma model with 
kT = 1 keV, N$_{\rm H}$ = 3 x 10$^{22}$ cm$^{-2}$, and d = 4 kpc unless otherwise noted,
and have typical uncertainties $\pm$0.4 dex. 
Candidate identifications lie within 1$''$ of the X-ray position and are from the  
{\it Hubble Space Telescope} Guide Star Catalog (GS) v2.2, the 2MASS (2M) data base, and 
archival {\em New Technology Telescope} (NT) images (J,H,K$_{s}$ bands). 
WR star identifications are from 
Table 2 of Negueruela \& Clark (2005) unless otherwise noted.
Exposure live times are obs1 (22-23 May 2005): 18,808 s, obs2 (18-19 June 2005): 38,473 s. } 
\tablenotetext{b}{Faint source, classified as a possible detection.}
\tablenotetext{c}{Tabulated data are from events collected in the first 18.8 ksec 
observation.}
\tablenotetext{d}{Source was detected in both observations. Tabulated data are from events 
collected in the second 38.5 ksec observation.}
\tablenotetext{e}{W9 is listed as source 9 in Table 1 of Clark et al. (2005) who classify 
it as sgB[e]. }
\tablenotetext{f}{Source 3 in Groh et al. (2006). }
\tablenotetext{g}{~L$_{x}$ is from spectral fit.}
\tablenotetext{h}{~Low level variability may be present in source 7 (WR-B) during the second observation, but a
                  K-S test gives P$_{c}$ = 0.86 for the first observation. The mean count
                  rates in both observations were the same to within the uncertainties.}
\end{deluxetable}

\clearpage

\clearpage

\begin{figure}
\figurenum{1}
\epsscale{1.0}
\includegraphics*[width=12.5cm,angle=0]{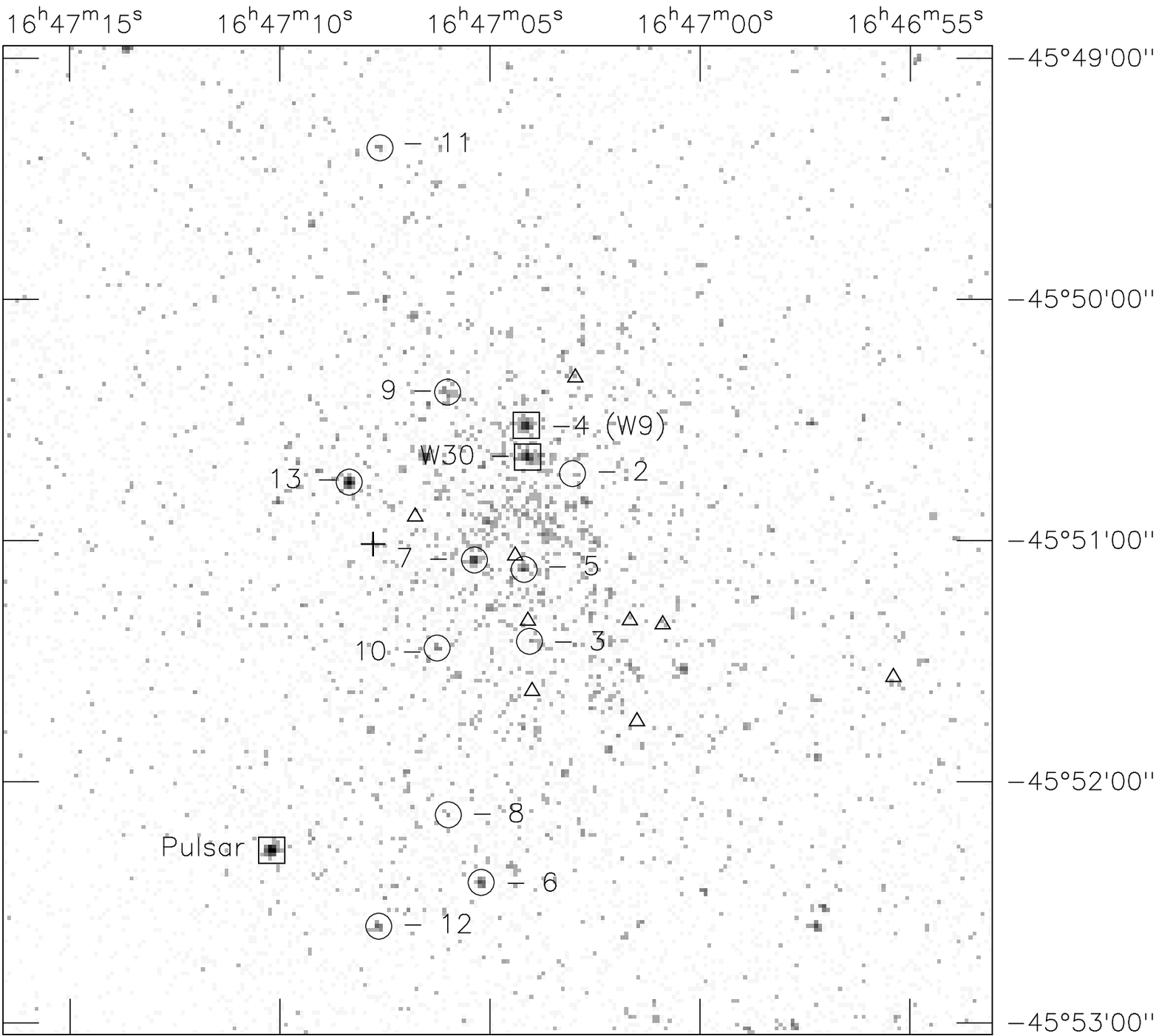}
\caption{
         Chandra ACIS-S image (0.3 - 7 keV) of the central region 
         of Wd 1 obtained on 18-19 June 2005 (38.5 ksec). 
         The image has a logarithmic stretch and is rebinned
         by a factor of two to a pixel size of 0.984$''$. A plus
         sign ($+$) marks the {\em Chandra} aimpoint. Circles enclose 
         X-ray detected WR stars (Table 1).
         Source 1 (WR-N) lies to the south and is not shown. 
         Triangles mark positions of undetected WR stars. 
         Squares enclose the bright X-ray sources
         W9 (B[e]sg), W30 (OB), and a newly-discovered X-ray
         pulsar.  Coordinates are J2000.
}

\end{figure}

\clearpage

\begin{figure}
\figurenum{2}
\epsscale{1.0}
\plotone{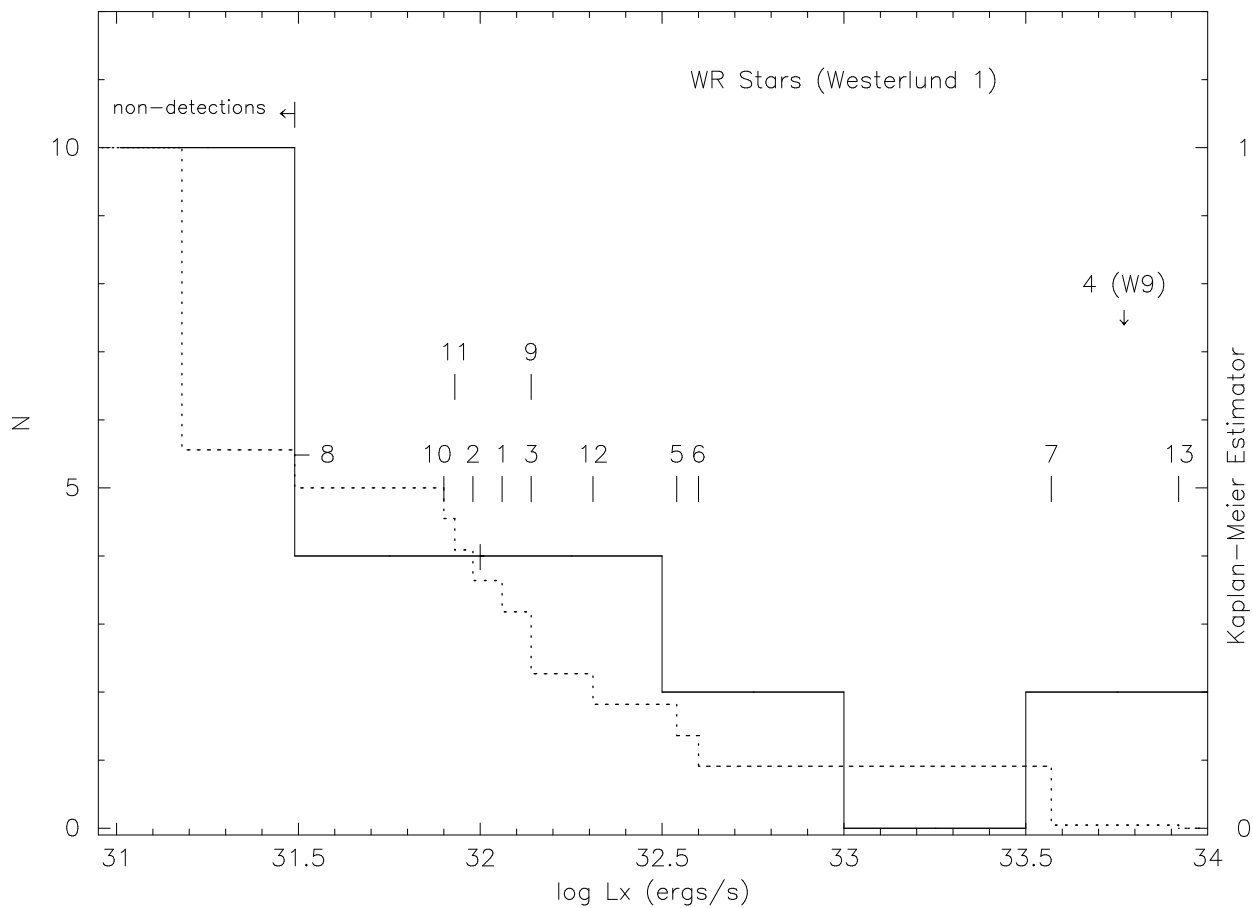}
\caption{
         Number of WR stars (N) versus unabsorbed X-ray luminosity
         L$_{x}$ (0.3 - 7 keV) in Wd 1, assuming d = 4 kpc 
         (B05). Source numbers correspond
         to Table 1. W9 is also shown for comparison. For brighter
         detections (4, 7, 13), L$_{x}$ was determined from spectral
         fits (Secs. 3.3, 4.0). For fainter detections and non-detections,
         L$_{x}$ was estimated from the observed count rate (or upper 
         limit) and the PIMMS simulator 
         (http://asc.harvard.edu/toolkit/pimms.jsp)
         assuming a 1T thermal plasma model with kT = 1 keV and
         N$_{\rm H}$ = 3$\times$ 10$^{22}$ cm$^{-2}$. The dotted
         line shows the Kaplan-Meier estimator for all 22 WR stars,
         taking upper limits into account.          
}
\end{figure}

\clearpage

\begin{figure}
\figurenum{3}
\epsscale{1.0}
\plotone{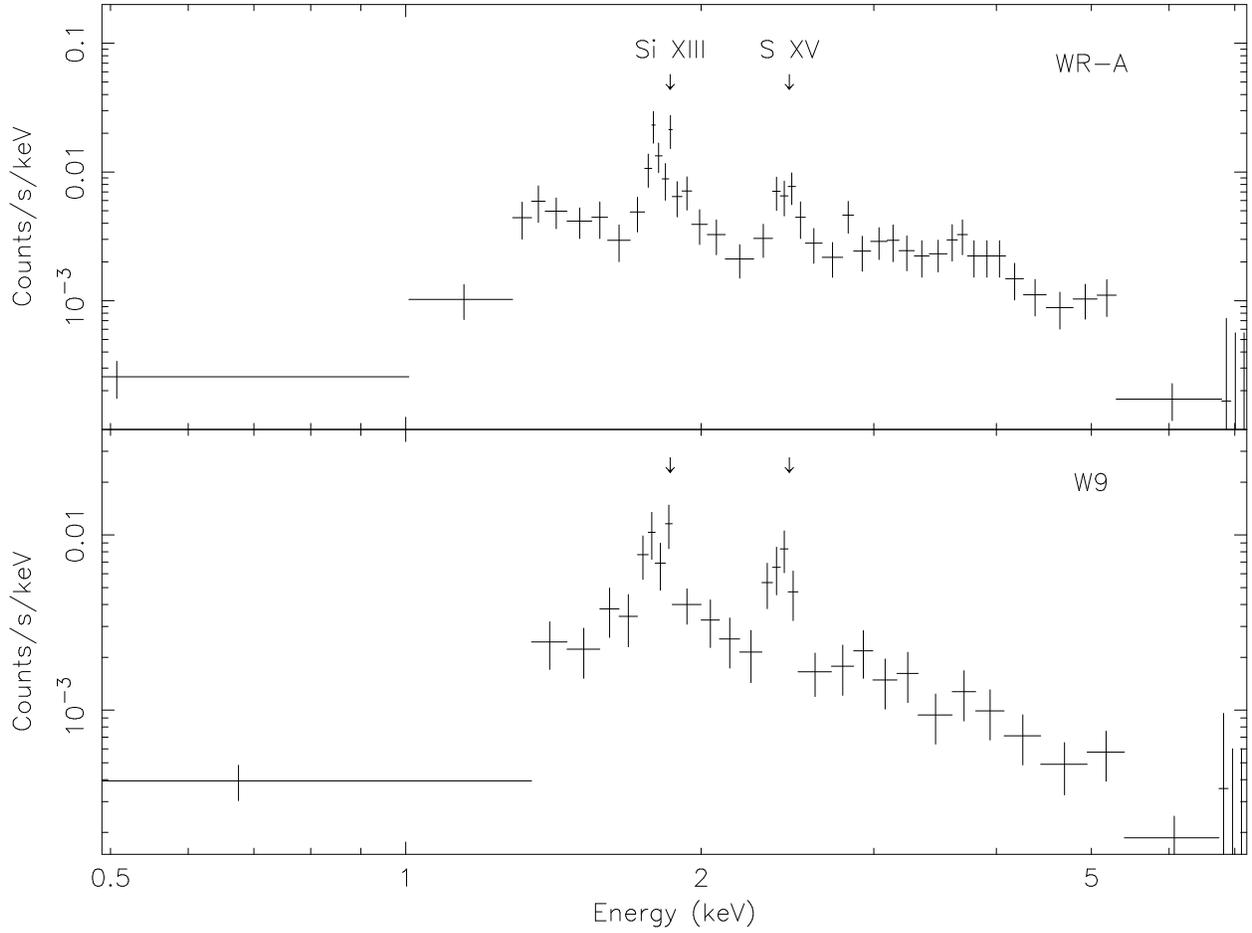}
\caption{
         Background subtracted {\em Chandra} ACIS-S 
         spectra of the WN star WR-A
         and the emission-line star W9. Spectra are
         from the 38.5 ksec observation on 18-19 June 2005
         and are binned to a minimum
         of 10 counts per bin.
}
\end{figure}

\clearpage

\end{document}